\documentclass[pop,fleqn]{w-art}
\usepackage{times}
\usepackage{w-thm}
\usepackage[]{graphicx}

\begin{document}

\DOIsuffix{theDOIsuffix}
\Volume{51}
\Issue{1}
\Month{09}
\Year{2005}
\pagespan{3}{}
\Receiveddate{15 September 2005}


\title[Creating and probing]{Creating and probing long-range order in atomic clouds}

\author[C. von Cube]{C. von Cube\inst{1}}
\address[\inst{1}]{Physikalisches Institut, Eberhard-Karls-Universit\"at T\"ubingen, 
\\Auf der Morgenstelle 14, D-72076 T\"ubingen, Germany}
\author[S. Slama]{S. Slama\inst{1}}
\author[M. Kohler]{M. Kohler\inst{1}}
\author[C. Zimmermann]{C. Zimmermann\inst{1}}
\author[Ph.W. Courteille]{Ph.W. Courteille\inst{1}\footnote{
	Corresponding author\quad E-mail:~\textsf{courteille@pit.physik.uni-tuebingen.de}, 
	Phone: +49\,29\,76\,279, 
	Fax: +49\,29\,58\,29}}

\begin{abstract}
Ultracold atoms interacting with the optical modes of a high-$Q$ optical ring cavity can synchronize their motion. The collective 
behavior makes the system interesting for quantum computing applications. This paper is devoted to the study of the collective coupling. 
We report on the first observation of a collective dynamics and on the realization of a laser, the gain mechanism of which is based on 
collective atomic recoil. We show that, if the atoms are subject to a friction force, starting from an unordered distribution they 
spontaneously form a moving density grating. Furthermore, we demonstrate that a 1D atomic density grating can be probed via Bragg 
scattering. By heterodyning the Bragg-reflected light with a reference beam, we obtain detailed information on phase shifts induced by 
the Bragg scattering process. 
\end{abstract}

\maketitle

\tableofcontents

\section{Introduction}
\label{SecIntroduction}

Ideally, a scalable quantum computer consists of registers of globally coupled two-level systems called \emph{qubits}. To match the 
requirement of individual addressability of the qubits with the necessity of long-range correlations a periodic ordering of the qubits 
seems desirable. To name only two examples out of a huge number of proposals for quantum computers: Chains of ions confined in Paul-traps 
\cite{Cirac95b,Nagerl00} can be coupled by electrostatic repulsion through a common mode of their center-of-mass oscillation. Arrays of 
neutral heteronuclear molecules suspended in free space are coupled through long-range dipole-dipole interactions \cite{DeMille02}. 

A proposal which is very similar to the ion chain idea consists in confining neutral atoms in the anti\-nodes of a standing light wave 
formed by the two counterpropagating modes of a high-finesse ring cavity \cite{Hemmerich99}. The atoms are periodically arranged in a 
one-dimensional optical lattice. The isolation of single atoms in standing waves and even their controlled shifting has impressively 
been demonstrated \cite{Alt03}. If several such atoms are confined in different antinodes, they can individually be addressed by probe 
beams. In fact, optical lattices draw an ever increasing attention, in particular in combination with quantum degenerate gases. 
E.g.~Mott-insulators \cite{Greiner02} may constitute good supports for quantum gates \cite{Mandel04}, and single atom transistors in 
optical lattices have been proposed \cite{Micheli04}. 

However, ring cavities provide an even more important advantage, because they combine the scalability of optical lattices with a 
strictly uniform coupling and a strength which \emph{increases} with the number of atoms in contrast to the situation in ion chains: The 
atoms all couple to the same light fields by scattering photons from one cavity mode into the counterpropagating one and receiving in 
return twice the photonic recoil. We have provided the proof that a collective dynamics can be driven in high-finesse ring cavities in 
recent experiments \cite{Kruse03b}. However these experiments deal with millions of thermal atoms, which is of course incompatible with 
individual addressability, and further research is needed to design an operating ring cavity quantum computer. 

\bigskip

In this paper we present our research project, which has the goal to experimentally study collective dynamics in high-finesse ring 
cavities. A priori it is not so clear, that such a dynamics can take place, because perturbative effects known from laser gyroscopes may 
lock the relative phase of the counterpropagating modes and freeze the essential degree of freedom needed to couple the atomic motions. 
We found however that under certain circumstances a collective instability leads to self-amplification and exponential gain of a light 
field on one hand and to atomic bunching on the other. A careful study of our system enabled us to directly relate the observations 
presented in Ref.~\cite{Kruse03b} to a phenomenon called the \emph{collective atomic recoil laser} (CARL) postulated 10 years ago 
\cite{Bonifacio04}. 

In the absence of dissipation for the motional degrees of freedom, the gain must remain transient. Consequently, we have observed a 
burst of light reverse to the pump beam direction. The light frequency increases in time as its amplitude vanishes. This burst is 
accompanied by a continuous acceleration of atoms. By introducing dissipation for their kinetic energy, we obtained a steady-state 
operation of the emitted light at a self-determined frequency. When the energy dissipation (or cooling) mechanism is limited to 
finite temperatures by some diffusion process, the appearance of a threshold is expected. We report here the observation of such a 
threshold: A minimum pump power is indeed necessary to start CARL action \cite{Cube04}. The threshold behavior of the cavity fields goes 
along with a phase transition in the atomic density distribution. 

\bigskip

This paper is divided into three parts. In section~\ref{SecCollective}, we outline the experimental apparatus and the sequence of our 
measurements. We emphasize that we have access to all relevant degrees of freedom characterizing our system: We monitor the dynamics of 
the cavity field by beating the field modes, record pictures of the atomic density distribution and probe the atomic velocity 
distribution by RIR spectroscopy. We present our observations and discuss the signatures of CARL in each of the monitored signals. We 
show that CARL is a transient phenomenon in the absence of dissipation for the external degrees of freedom. In contrast, in the presence 
of a friction force (experimentally realized by an optical molasses) a steady state is reached. In section~\ref{SecKuramoto} we draft a 
theoretical model that we compare to our observations. An analytic treatment for the ideal case of perfect atomic bunching enables us to 
find compact solutions for the probe field intensity and frequency. Numerical simulations of Langevin (or alternatively Fokker-Planck) 
equations at finite temperature show, that dissipation gives rise to a threshold behavior, which draws an analogy between CARL and 
conventional lasers. They also reveal that driven by a dissipative force, homogeneous atomic clouds can show collective instabilities 
spontaneously producing long-range order, atomic bunching and a thermodynamic phase transition reminiscent to self-organization 
mechanisms in networks of coupled oscillators. Section~\ref{SecBragg} focusses on the detection of long-range order via Bragg 
scattering. This part of the work is obviously motivated by our desire to unveil the most direct signature of CARL, which is atomic 
bunching. But we will see that the Bragg scattering method has also features, which are interesting in their own rights.

\section{Collective coupling}
\label{SecCollective}

A consequence of the development of new powerful techniques for optical cooling and trapping in the past decades is the ever increasing 
importance of the role assigned to the center-of-mass degrees of freedom in the interaction of matter with light. Prominent examples of 
how the motion of atoms can influence the scattering of light are the recoil-induced resonances (RIR) and the CARL. The RIR are observed 
\cite{Guo92,Courtois94} as Raman transitions between momentum states. Two laser beams having different frequencies and crossing each 
other under a finite angle give rise to a moving standing light wave. Atoms moving synchronously to this wave can be scattered to other 
momentum states. The scattering process is monitored via power variations in the Raman beams. The spectroscopy of RIR is today routinely 
used to characterize the velocity-distribution of atomic samples \cite{Kruse03}. 

The CARL has been conceived as a collection of two-level atoms driven by a single-mode pump field \cite{Bonifacio94}. Under appropriate 
conditions, atomic density fluctuations (together with excitation and polarization fluctuations) induce a small amount of backscattering 
which interferes with the pump field and creates a weak traveling modulation wave. This wave, in turn, generates a weak density wave 
which, for appropriate values of the parameters, radiates and strengthens the backward scattered field in an avalanche process. The 
signature of CARL (and this distinguishes it from RIR) is thus an \emph{exponential growth} of the backward field. In fact, atomic 
bunching and probe gain can arise \emph{spontaneously} from fluctuations with no seed field applied \cite{Bonifacio94}. An essential 
feature of the CARL model is the dynamical role played by the atomic motion. It must be included in the theoretical description to give 
proper account of recoil effects, which eventually are responsible for producing the organized atomic density modulation, or density 
grating, which is at the heart of the CARL process. 

\bigskip

The CARL effect should emerge most clearly in cold atomic clouds in a collisionless environment \cite{Berman99}. Furthermore, large 
detunings of the lasers far outside the Doppler-broadened profiles of the atomic resonances are preferable. In this regime, excited 
states can safely be adiabatically eliminated, and effects from atomic polarization gratings \emph{not} based on density variations, 
are avoided. Finally, to emphasize the role of exponential gain responsible for self-bunching, i.e.~spontaneous formation and growth 
of a density grating, it is desirable not to seed the probe. The observation of a probe beam is then a clear indication for CARL 
\cite{Kruse03b}. Earlier attempts to observe CARL action have been undertaken in hot atomic vapors with near-resonant laser beams 
\cite{Lippi96b,Hemmer96}. They have led to the identification of a reverse field with some of the expected characteristics. However, the 
gain observed in the reverse field can have other sources \cite{BrownWJ97,BrownWJ97b}, which are not necessarily related to atomic 
recoil. 

The best way to satisfy the specified conditions is to recycle the probe light in a very high finesse ring cavity. This permits to 
realize strong atom-field coupling forces very far from atomic resonances, i.e.~in a regime where the atomic excitation plays no role. 
The large power enhancement inside a high-finesse cavity allows to confine the atoms inside the cavity mode-volume and thus to have 
quasi unlimited interaction times. The atomic cloud can be as cold as $100~\mu$K. The densities are generally low enough to permit only 
few collisions per second. Far from resonance spontaneous scattering and thus radiation pressure is in most cases negligible. 

Apart from ours, other experiments with cold atoms have recently found signatures for collective dynamics. Optical bistability has been 
observed in a high-finesse ring cavity \cite{Nagorny03b,Elsasser03}. Spatial self-organization induced by collective friction force has 
been observed in a linear low-finesse cavity \cite{Black03}. In both cases the underlying dynamics is however different from CARL 
\cite{Domokos03}.

\subsection{Experimental setup}
\label{SecCollectiveExperiment}

Ring cavities are particular in the following sense. The optical dipole force exerted by the light fields on atoms can be understood in 
terms of momentum transfer by coherent redistribution (via Rayleigh scattering) of photons between optical modes. In multimode 
configurations, e.g.~counterpropagating modes of a ring cavity, the photon pumping can occur between \emph{different} modes having 
independent photon budgets \cite{Gangl00b}. As a consequence, the scattering process conserves the total momentum of the atoms and 
photons, which is not the case for linear cavities, where backscattered photons end up in the same cavity mode. This implies that the 
atoms exert a noticeable backaction on the optical fields and that the atomic dynamics can be monitored as an intensity 
\cite{Kozuma96,Raithel98} or a phase imbalance between the modes \cite{Kruse03b}. Consequently, in a ring cavity the scattering of a 
photon between the counterpropagating modes slightly shifts the phase of the standing wave.\footnote{
	In the absence of spatially fixed backscattering objects the phase of the standing wave is free, in contrast to linear cavities 
	where the phase is fixed by boundary conditions at the mirror surfaces.}
This shift, being strongly enhanced by a long cavity life time, is sensed by all atoms trapped in the standing wave. Consequently, the 
simultaneous interaction of the atoms with the two field modes couples the motion of all atoms \cite{Hemmerich99,Gangl00}.

\subsubsection{Ring cavity}
\label{SecCollectiveExperimentRing}

\begin{figure}[h]
		\includegraphics[width=7.5cm]{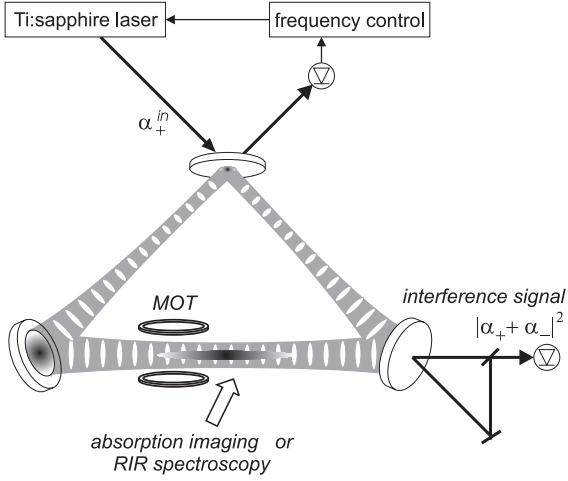}\caption{
		Scheme of the experimental setup. A titanium-sapphire laser is locked to one of the two counterpropagating modes ($\alpha_+$) of 
		a ring	cavity. The atomic cloud is located in the center of the magneto-optical trap (MOT) close to the free-space waist of the 
		cavity mode. We observe the evolution of the interference signal between the two light fields leaking through one of the cavity 
		mirrors, record the spatial evolution of the atoms via absorption imaging, and measure their velocity distribution by spectroscopy 
		of recoil-induced resonances (RIR).}
		\label{CollectiveSetup}
		\end{figure}

The ring cavity which is at the heart of our experiment (see Fig.~\ref{CollectiveSetup}) has a round trip length of $L=85$~mm and beam 
waists in horizontal and vertical direction at the location of the MOT of $w_v=129~\mu$m and $w_h=124~\mu$m, respectively 
\cite{Kruse03}. This corresponds to the cavity mode volume $V_{mode}=\frac{\pi}{2}Lw_vw_h=2$~mm$^3$. For (linear) $s$-polarization the 
two curved high reflecting mirrors have a transmission of $2\times10^{-6}$, while the plane input coupler has a transmission of 
$27\times10^{-6}$. Depositions of rubidium on the mirror surfaces actually reduce the finesse to $F=80000$, measured by cavity ring-down. 
This corresponds to an amplitude decay rate of $\kappa=\pi\delta/F=2\pi\times 22$~kHz, where $\delta=c/L$ denotes the free spectral 
range. For $p$-polarized light we found $F=2500$. 

The optical layout of our experiment is shown in Fig.~\ref{CollectiveSetup}. A titanium-sapphire laser is double-passed through an 
acousto-optic modulator which shifts the laser frequency. The light beam is then mode-matched to one of the two counterpropagating 
cavity modes. The reflected light bears information about phase deviations between the laser and the intracavity light field. The 
light reflected from the cavity is fed back via a Pound-Drever-Hall type servo control to phase correcting devices. The phase 
corrections are made via a piezo transducer mounted to the titanium-sapphire laser cavity and via the acousto-optic modulator. The 
frequency deviations of the stabilized laser from the ring cavity resonance are much less than the linewidth of the ring cavity. Because 
the cavity mirrors have very different reflectivities for $s$- and $p$-polarization, we can switch from high to low finesse by just 
rotating the linear polarization of the injected laser beam. This tool is useful for tests comparing strong and weak coupling situations. 

From the electric field per photon $\mathcal{E}_1=\sqrt{\hbar\omega/2\varepsilon_0V_{mode}}$ and the dipole moment 
$d\approx\sqrt{3\pi\varepsilon_0\hbar\Gamma/k^3}$ \cite{Grimm00}, we calculate the atom-field coupling constant (or single-photon Rabi 
frequency) to be $g_0=d\mathcal{E}_1/\hbar\approx2\pi\times100~$kHz, where $\Gamma=2\pi\times6~$MHz is the natural linewidth of the $D_2$ 
line of $^{85}$Rb. Therefore the coupling strength is much larger than the cavity decay rate, but still much smaller than the natural 
linewidth, $\kappa\ll g_0\ll\Gamma$. 

In the following, we label the modes by their complex field amplitudes scaled to the field per photon, so that 
$I_{\pm}=2\varepsilon_0 c\mathcal{E}_1^2|\alpha_{\pm}|^2$ is the intracavity light intensity in the respective mode $\alpha_{\pm}$. The 
mode $\alpha_+$ is pumped with the rate $\eta_+\equiv\sqrt{\delta\kappa}~\alpha_+^{in}$, where $\alpha_+^{in}$ is the field amplitude of 
the incoupled laser beam (under the premise of perfect phase and impedance matching). On resonance and in the absence of any losses 
other then those included in $\kappa$, the cavity dramatically enhances the incident light power by 
$\alpha_+=\sqrt{\delta/\kappa}~\alpha_+^{in}$ \cite{Elsasser03}. The counterpropagating mode $\alpha_-$ is populated out of $\alpha_+$ by 
backscattering from atoms located inside the mode volume or from mirror imperfections. The light power outcoupled through a mirror with 
reflectivity $T$ is related to the intracavity power $P_{\pm}=\tfrac{\pi}{2}w_h w_v I_{\pm}$ via 
$P_{\pm}^{out}=TP_{\pm}=T\hbar\omega\delta~|\alpha_{\pm}|^2$.

\subsubsection{Dipole trap for $^{85}$Rb}
\label{SecCollectiveExperimentDipole}

The ring cavity is located inside an ultrahigh vacuum recipient. We produce rubidium-85 atoms by a dispenser and collect them with a 
standard magneto-optical trap before transferring them into a TEM$_{00}$ mode of the ring cavity field dipole potential. We typically 
load several million atoms into the dipole trap at temperatures of a few $100~\mu$K and atomic peak densities of about 
$2\times10^9$~cm$^{-3}$. Density and temperature are monitored by time-of-flight absorption imaging. 

The dipole force acting on the atoms and allowing for their confinement originates from intensity gradients in the cavity field. Let us 
assume that the field frequency $\omega$ is red-detuned from an atomic resonance $\omega_a$, $\Delta_a\equiv\omega-\omega_a<0$. Radial 
confinement is then ensured by the transverse Gaussian profile of the TEM$_{00}$ mode. Axially, the atoms tend to concentrate around the 
free space waist of the cavity mode. The optical potential of the dipole force trap is, for a far detuned laser frequency, 
$|\Delta_a|\gg\Gamma$, to a good approximation given by $\phi(\mathbf{r})=\hbar\Omega (\mathbf{r})^2/4\Delta_a$, where 
$\Omega^2(\mathbf{r})=4g^2 (\mathbf{r})|\alpha_+e^{ikz}+\alpha_-e^{-ikz}|^2$ is the local Rabi frequency. Obviously, the atom-field 
coupling strength $g(\mathbf{r})$, also called the cavity mode function, becomes inhomogeneous. $g_0=g(0)$ is the one-photon Rabi 
frequency at the center of the trap, whose expression is given above. 

Since rubidium exhibits a fine structure, we must consider the contributions of the $D_1$ and the $D_2$ line to the optical potential 
\cite{Grimm00}: 
	\begin{align}
	\phi(\mathbf{r}) & =\hbar U(\mathbf{r})|\alpha_+e^{ikz}+\alpha_-e^{-ikz}|^2~,\tag{1}\nonumber\\
	\text{where \ \ \ \ } U(\mathbf{r}) & \equiv g(\mathbf{r})^2\left(\frac{1}{\Delta_{D1}}+\frac{2}{\Delta_{D2}}\right)~.\nonumber
	\end{align}
In a running wave laser beam, $\alpha_-=0$, close to the center of the trap (i.e.~on the cavity axis, $r\ll w_0\equiv\sqrt{w_vw_h}$, and 
within the Rayleigh length, $z\ll\frac{1}{2}kw_0^2$) the normalized mode function can be approximated by a harmonic potential 
$U(\mathbf{r})\approx U_0 \left[1-\left(2z/kw_0^2\right)^2-2\left(r/w_0\right)^2\right]$. For a typical intracavity light power of 
$P=10~$W and a frequency detuning from the nearest atomic resonance of typically $\Delta_{D1}=-2\pi\times1~$THz, we calculate an optical 
potential depth of $\phi_0\approx k_B\times 1.4~$mK. The radial and axial secular frequencies of the trap are 
$\omega_r\approx 2\pi\times1~$kHz and $\omega_z\approx 2\pi\times2~$Hz. An important quantity characterizing the coupling strength and 
used throughout this paper, the \emph{one-photon light shift}, is typically on the order of $U_0\equiv U(0)\approx -0.01$. 

Not too close to the resonance, the spontaneous scattering rate out of one beam is well approximated by 
	\begin{align}
	\gamma_{scat}(\mathbf{r}) & =\gamma(\mathbf{r})|\alpha_+|^2~,\tag{2}\nonumber\\
	\text{where \ \ \ \ } \gamma(\mathbf{r}) & \equiv g(\mathbf{r})^2\Gamma\left(\frac{1}{\Delta_{D1}^2}+\frac{2}{\Delta_{D2}^2}
		\right).\nonumber
	\end{align}
It decreases faster with detuning than the potential depth, so that for a chosen trap depth it is advantageous to work at higher 
detunings and higher laser intensities in order to avoid excessive heating by spontaneous scattering processes. For the parameters given 
above, we expect $\gamma_0\approx 6\times 10^{-8}~\text{s}^{-1}$ and $\gamma_{scat,0}\approx 600~\text{s}^{-1}$. In this context the 
high-finesse optical cavity has the practical advantage that its large intensity enhancement \cite{Mosk01} ensures a strong atom-field 
coupling even at laser detunings as large as $1..10$~nm. 

\bigskip

Collisions are predicted to hamper the CARL process in buffer gas cells \cite{Perrin02,Javaloyes03}, because they remix the atomic 
distributions faster than bunching through CARL dynamics. In contrast, for cold atoms the collision rate is generally quite small. At 
our typical densities and temperatures, we estimate a collision rate of $\gamma_{coll}=n\sigma \overline{v}/2\approx 1$~min$^{-1}$, 
where $\overline{v}=\sqrt{k_B T/m}$ the average velocity and $\sigma\approx 10^{-12}$~cm$^2$ the elastic collision cross section 
\cite{Burke98}. This shows that on the time scale of our experiments (typically ms) the impact of collisions can be discarded.

\subsubsection{Optical molasses}
\label{SecCollectiveExperimentMolasses}

Some of our experiments are carried out in the presence of a so-called optical molasses. Technically, to create a molasses we use the 
laser beams of the magneto-optical trap, but without magnetic field gradients. Thus the molasses consists of three orthogonal pairs of 
counterpropagating $\sigma^\pm$ polarized laser beams, which are tuned a few natural linewidths $\Gamma$ below the $D_2$ resonance. They 
traverse the dipole trap at angles of either $45^\circ$ or $90^\circ$. The molasses light is linearly polarized at any location with a 
polarization vector winding about the molasses beam directions. Through these polarization gradients the atomic cloud is cooled well 
below the Doppler limit. The atomic motion inside an optical molasses is well described as being subject to a velocity-dependent 
friction force: The atoms are slowed down and move like in a viscous fluid, but there is no restoring force; the atoms are not trapped. 

The friction force sensitively depends on the detuning of the molasses beams from resonance. Unfortunately, the dipole trap produces a 
light shift of the electronic states, which are coupled by the $D_1$ and $D_2$ transitions. The light shift increases the effective 
detuning of the molasses lasers by an amount equivalent to $U(\textbf{r})$. The effective detuning thus gets inhomogeneous.\footnote{
	As long as the dipole trap laser detuning greatly exceeds the hyperfine splittings of the ground and excited states, at least the 
	optical potential is the same for all hyperfine and magnetic sublevels \cite{Grimm00}.} 

Another potential problem is the following: In the presence of magnetic fields the atoms are optically pumped towards fully stretched 
Zeeman states. Once the atoms are doubly spin-polarized, they preferentially interact with \emph{only one} of the counterpropagating 
molasses beams, which leads to a radiation pressure imbalance. The resulting atomic acceleration interferes with signatures of CARL 
dynamics (see next sections). In practice, we take care to accurately compensate the magnetic fields. 

Our main purpose in using optical molasses is to introduce friction to the system. However, the scattering of light by the molasses also 
leads to diffusion in momentum space, which limits the temperature to which the atoms are cooled. We will see in 
Sec.~\ref{SecKuramotoTrajectories} that the interplay of friction and diffusion has a large impact on the dynamics of the system.

\subsection{Signatures of collective atomic recoil lasing}
\label{SecCollectiveSignatures}

A typical experimental sequence goes as follows \cite{Kruse03b}: After the dipole trap has been filled with atoms, a time of about 
$100~$ms is waited until the atoms have found their thermal equilibrium and form a homogeneous cloud. The CARL process is started by 
\emph{suddenly switching on the optical molasses}. The molasses beams are extinguished again after some 100~ms.

The observables of our system are the field amplitudes $\alpha_\pm$ and the atomic coordinates $x$ and $p$. In order to clarify the 
interplay between these degrees of freedom, we measure the response of the coupled atom-cavity system to switching on and off the 
molasses. Three signals are monitored: First of all, we record the interference signal obtained from a frequency beat of the two 
counterpropagating ring cavity modes. Second, time-of-flight absorption images allow us to detect spatial displacements of the atomic 
cloud or to measure its temperature. Finally, we record RIR spectra of the atomic momentum distribution thus obtaining complementary 
information on the atomic cloud's state of motion.

\subsubsection{Beat note of field modes}
\label{SecCollectiveBeatnote}

The phase dynamics of two counterpropagating cavity modes is monitored as a beat signal between the two beams, which are outcoupled at 
one of the high-reflecting cavity mirrors (see Fig.~\ref{CollectiveSetup}) and phase-matched on a photodetector. Any frequency 
difference between pumped mode ($\omega_+$) and reverse (probe) mode ($\omega_-$) gives rise to a propagation of the standing wave 
inside the ring cavity, which translates into an amplitude variation of the beat signal, 
	\begin{equation}
	P_{beat}=T\hbar\omega\delta~|\alpha_++\alpha_-|^2~.\tag{3}\nonumber
	\end{equation}
The beat signal oscillates with the frequency $\Delta\omega\equiv\omega_+-\omega_-$. 

\begin{figure}[h]
		\includegraphics[width=6cm]{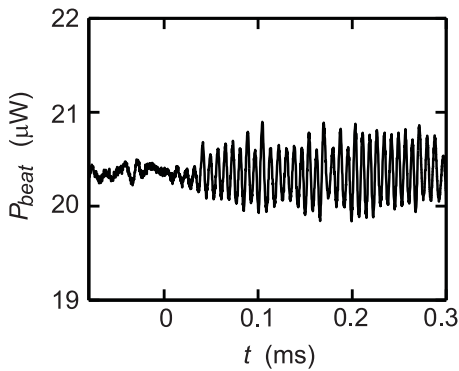}\caption{
		Recorded time evolution of the beat signal between the two cavity modes \cite{Kruse03b}. At time $t=0~$s the molasses beams have 
		been switched on.}\label{CollectiveObservation}
		\end{figure}

Fig.~\ref{CollectiveObservation} shows the time evolution of the beat signal. Initially, apart from a certain amount of noise, there is 
no discernable signal. But as soon as the molasses is irradiated at time $t=0~$s, strong oscillations appear with a fixed frequency. 
They persist for more than 100~ms. The oscillations \emph{do not arise}, when the cavity is set up for low finesse. 

To understand this observation, we consider an atomic cloud irradiated by a pump laser. As long as the cloud is homogeneous, the pump 
light is not coherently backscattered, because the backscattered photons have random phases and interfere destructively. Tiny density 
fluctuations may however give rise to a small amount of net backscattering. The backscattering atoms are accelerated due to photonic 
recoil, and the backscattered light is red-detuned because of the Doppler effect. Together with the pump light the backscattered probe 
generates by interference a weak modulation of the intracavity light intensity, i.e.~a standing wave fraction, which propagates in the 
same direction as the pump beam. If no other forces are applied, the accelerated atoms spread out in space and merge with the bulk 
of the homogeneous cloud; the fluctuation disappears and the probe decays. If however an optical molasses is irradiated, the acceleration 
is balanced by a friction force. The atoms now have time enough to probe the dipole potential of the standing wave and be pulled towards 
the potential valleys. Therefore they arrange themselves into a periodic pattern, which dramatically increases the backscattering 
efficiency due to Bragg reflection. The increased standing wave contrast in turn amplifies the tendency of the atoms to self-organize, 
and so on. Positive feedback results in exponential gain, only limited by the detuning of the probe mode from the cavity resonance 
(typically a few \% of the pump power). The propagating standing wave drags along the atoms, which in turn haul the standing wave. The 
propagation velocity corresponds to an equilibrium between the acceleration force exerted by the coherent backscattering and the 
velocity-dependent friction force exerted by the molasses. The atoms behave like \emph{surfing on a self-generated standing light wave}. 

\begin{figure}[h]
		\centerline{\scalebox{0.88}{\includegraphics{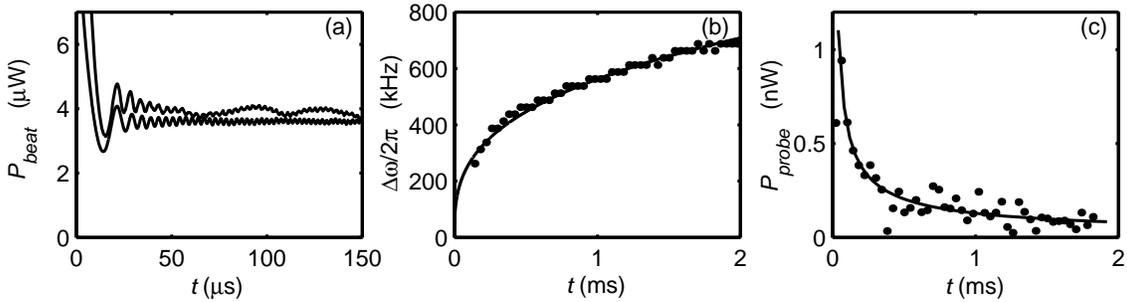}}}\caption{
		Acceleration of the standing wave as the molasses beams are interrupted. \textbf{(a)} Measured beat signal (upper curve) together 
		with a numerical simulation (lower curve). The instantaneous frequency difference between pump and probe is plotted in 
		\textbf{(b)} as a function of time. \textbf{(c)} Instantaneous fringe contrast, which is proportional to the probe beam intensity. 
		The solid lines represent calculations obtained from analytical formulae [Eqs.~(9) in Sec.~\ref{SecKuramotoAnalytic}] 
		\cite{Cube04}.}
		\label{CollectiveAccelerate}
		\end{figure}

As soon as the optical molasses is turned off, the dipole force has no counterbalance. The standing wave and the atoms accelerate each 
other and start to run without bounds. The frequency difference between the pump and the reverse lasing mode continuously increases. 
Consequently, the beat signal oscillates faster and faster, and the contrast of the interference fringes gradually fades out, because the 
frequency of the backscattered light deviates more and more from the cavity resonance [see Fig.~\ref{CollectiveAccelerate}(a)]. However 
fringes are still visible after more than $2$~ms, which corresponds to 600 cavity decay times. 

A Fourier analysis over consecutive time intervals of the beat signal $P_{beat}$ reveals the gradual frequency increase of the 
oscillations $\Delta\omega$, corresponding to an increasingly red-detuned probe beam [see Fig.~\ref{CollectiveAccelerate}(b)]. The 
decrease of the amplitude is visible in Fig.~\ref{CollectiveAccelerate}(c). In practice, we determine the probe beam power, 
$P_{probe}=T\hbar\omega\delta|\alpha_-|^2$, from the contrast of the beat signal (or the height of the peak in the Fourier spectrum), 
$\Delta P_{cont}\equiv \max{P_{beat}}-\min{P_{beat}}=4T\hbar\omega\delta|\alpha_+||\alpha_-|$, 
	\begin{equation}
	P_{probe}=\frac{\Delta P_{cont}^2}{16P_{pump}}~.\tag{4}\nonumber
	\end{equation}

\subsubsection{Spectra of recoil-induced resonances}
\label{SecCollectiveMomentum}

The above interpretation of the beat signal observation in terms of a propagating intracavity standing wave postulates an acceleration 
of the atomic cloud in the direction of the pump light. With the aim to provide evidence for an atomic motion, we map the atomic 
velocity distribution by RIR spectroscopy \cite{Kruse03,Meacher94}. Two laser beams in Raman configuration propagating within the ring 
cavity plane and enclosing a small angle $\theta=8^\circ$ are aligned nearly perpendicularly to the dipole trap. The Raman beams are 
tuned a few $100~$MHz blue to the $D_2$ line. 

\begin{figure}[h]
		\centerline{\scalebox{0.6}{\includegraphics{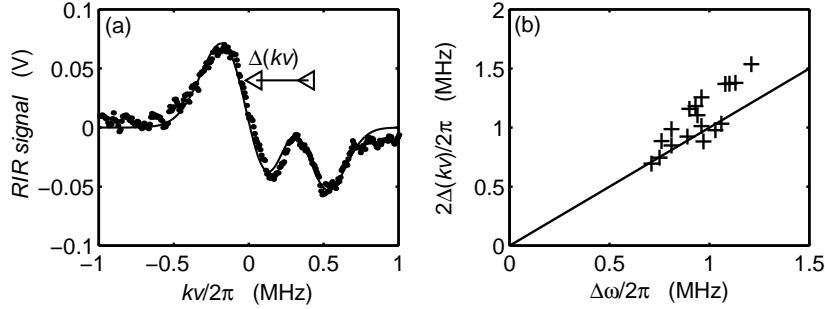}}}\caption{
		\textbf{(a)} RIR momentum spectra of a CARL-accelerated atomic cloud. Note that the velocity axis is rescaled in terms of the 
		Doppler shift occurring at the cavity field wavenumber $k$. The fitted curve assumes a superposition of two displaced Gaussian 
		derivatives. \textbf{(b)} Splitting $\Delta(kv)$ of the two Gaussians shown in (a) for various CARL acceleration times plotted as 
		a function of the beat frequency $\Delta\omega$ measured shortly before recording the RIR spectrum \cite{Courteille04}.}
		\label{CollectiveRir}
		\end{figure}

The RIR spectra are recorded a few ms after the molasses beams have been switched off and immediately after the pump laser has been 
interrupted, thus suddenly releasing the atoms from the dipole trap. The frequency of one Raman beam is frequency-ramped across a 
detuning range of $\Delta_{rir}/2\pi=-200..200~$kHz with respect to the other beam with a scan rate of $2~$kHz/$\mu$s. The power 
transmission of the fixed-frequency Raman beam through the atomic cloud is recorded on a photodetector. Fig.~\ref{CollectiveRir}(a) 
shows such a RIR spectrum. 

As pointed out in Ref.~\cite{Kruse03,Courteille04,Kruse04}, the net rate for scattering photons between the Raman beams, $W_{RIR}(v)$, is 
proportional to the number of atoms $N$, to the derivative of their velocity distribution $\partial\Pi(v)/\partial v$, and to a constant 
$C$ which depends on the intensities of the Raman beams. The velocity class of atoms contributing to the signal is selected via 
$v=\Delta_{rir}/q$, where $q=2k_{rir}\sin\tfrac{\theta}{2}$ and $k_{rir}$ is the wavenumber of the RIR beams. The lineshape recorded in 
Fig.~\ref{CollectiveRir}(a) suggests a bimodal velocity distribution for the CARL-accelerated cloud, which we model by a superposition of 
\emph{two distinct thermal ensembles} having different atom numbers $N_j$, different temperatures $T_j$, and different center-of-mass 
velocities $v_j$: 
	\begin{equation}
	W_{RIR}(v)=\sum_{j=1}^2\frac{CN_j}{\sigma_j\sqrt{2\pi}}~\frac{\partial}{\partial v}e^{-(kv-kv_j)^2/\sigma_j^2}~,\tag{5}\nonumber
	\end{equation}
where $\sigma_j^2=2k_BT_j~k^2/m$. From a fit of this formula to the RIR signal exhibited in Fig.~\ref{CollectiveRir}(a), we determine 
the relative atom number $N_2/N_1=0.9$, the temperatures $T_1=300~\mu$K and $T_2=200~\mu$K, and the relative Doppler shift 
$\Delta(kv)\equiv kv_2-kv_1=2\pi\times400~$kHz. The actual values depend on the experimental settings. In particular we find that the 
relative Doppler shift increases with the time delay between molasses switch-off and RIR scan. By plotting the relative Doppler shift 
against the instantaneous CARL frequency independently extracted from a simultaneously recorded beat signal [see 
Fig.~\ref{CollectiveAccelerate}], we obtain the curve in Fig.~\ref{CollectiveRir}(b). The data points correspond to different time 
delays. The equivalence $\Delta\omega=2\Delta(kv)$ nicely demonstrates that atoms are synchronously accelerated with the moving standing 
wave. On the other hand the bimodal feature of the velocity distribution shows, that a part of the cloud is apparently unaffected by the 
CARL dynamics.

\subsubsection{Atomic transport}
\label{SecCollectiveTransport}

\begin{figure}[h]
		\includegraphics[width=8cm]{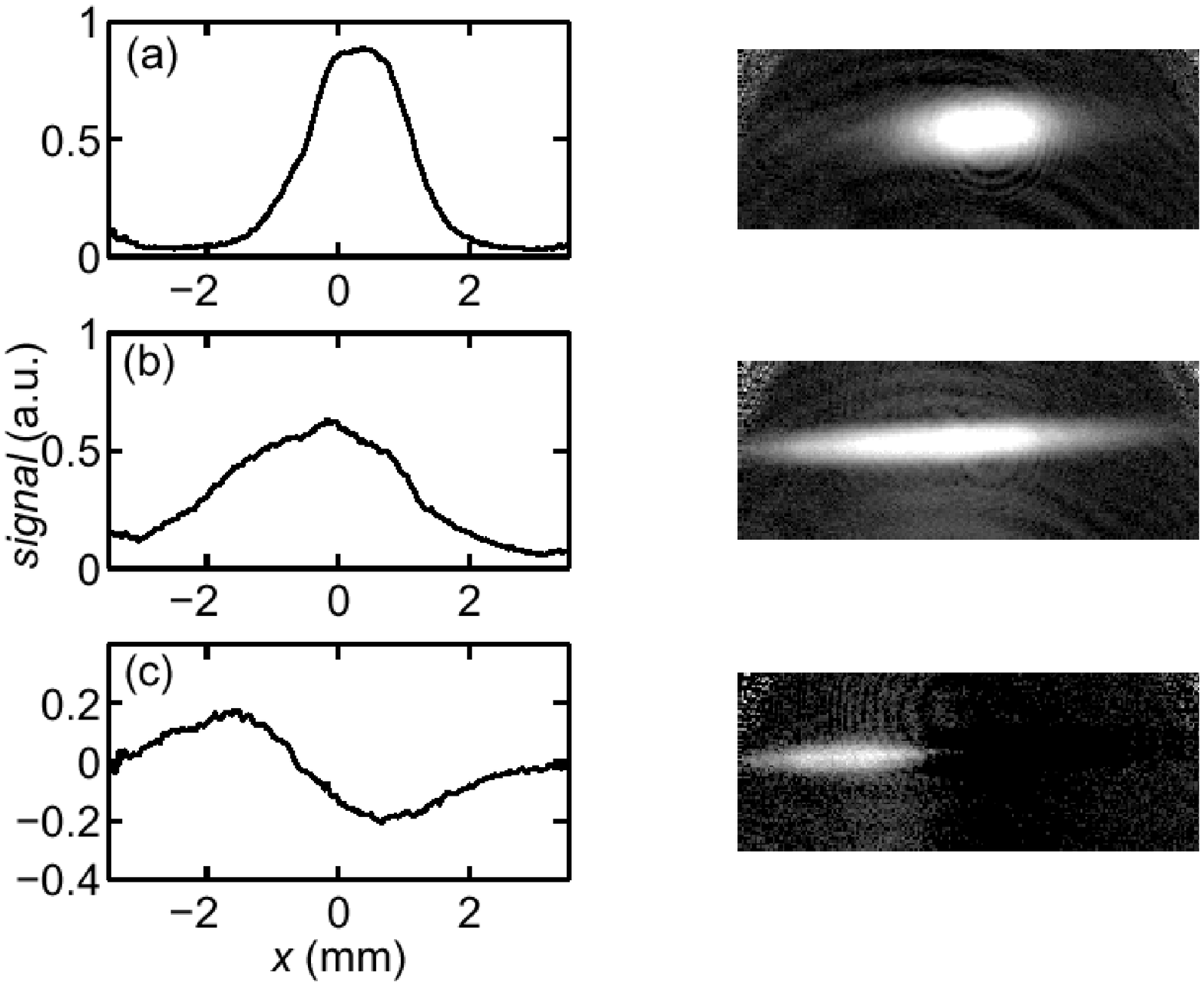}\caption{
		\textbf{(a)} Absorption images of a cloud of $6\times 10^6$ atoms recorded for high cavity finesse after $0~$ms and \textbf{(b)} 
		$6~$ms of CARL acceleration \cite{Kruse03b}. Image \textbf{(c)} emphasizes the CARL displacement by subtracting from image 
		\textbf{(b)} an absorption image taken with low cavity finesse after a $6~$ms CARL acceleration period. The intracavity power has 
		been adjusted to the same value as in the high-finesse case.}
		\label{CollectiveTof}
		\end{figure}

A Doppler-shift of $kv=2\pi\times500$~kHz corresponds to an atomic velocity of $40~$cm/s and leads to macroscopic displacements of the 
atoms on the order of several mm after a few ms. To observe this displacement, we take absorption pictures of the atomic cloud after a 
few ms of undamped CARL acceleration followed by a short 1~ms free expansion period.\footnote{ 
	For these measurements, the atomic bunching which precedes the CARL acceleration has \emph{not} been realized by an optical molasses. 
	Instead, the atoms were loaded into a standing wave dipole trap obtained by bidirectional pumping of the cavity \cite{Kruse03b}. This 
	method concentrates the atoms within a smaller segment along the cavity axis and makes the axial displacement due to CARL more 
	visible.} 
Fig.~\ref{CollectiveTof}(a) and (b) show pictures taken for CARL acceleration times of $0~$ and $6~$ms, respectively. After 6~ms a 
spatial shift of the cloud is clearly visible. 

In order to rule out that the atoms are simply shifted by radiation pressure, we repeat the measurement for weak coupling, i.e.~by 
operating the ring cavity at low finesse and rising the pump power, so that the intracavity power and consequently the radiation 
pressure are unchanged. Under these conditions we do not observe a shift of the atomic cloud for acceleration times below $10~$ms. 

The displacement can be estimated by integrating the formula~(9) in Sec.~\ref{SecKuramotoAnalytic}, which describes the time dependence 
of the Doppler shift: 
	\begin{equation}
	kx=\tfrac{3}{8}\Delta\omega t~.\tag{6}\nonumber
	\end{equation}
This yields $x\approx1.8$~mm, which agrees fairly well with the observation in Fig.~\ref{CollectiveTof}. A bimodal structure of the 
spatial distribution, as observed in the velocity distribution, is not discernible because of the large axial extend for the cloud.

\section{Creating long-range order}
\label{SecKuramoto}

\begin{figure}[h]
		\includegraphics[width=14.5cm]{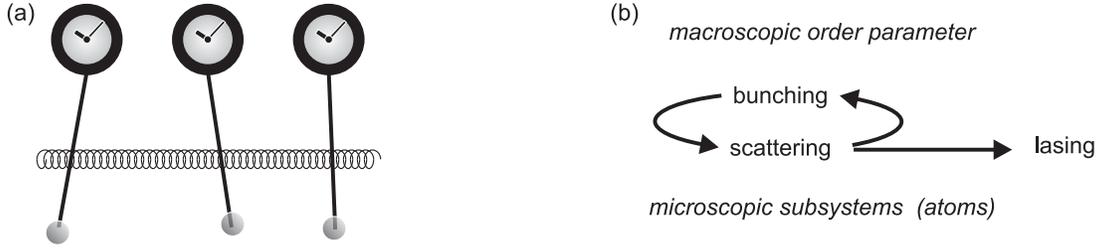}\caption{
		\textbf{(a)} Scheme of Huygens' coupled pendulum clocks. \textbf{(b)} Sketch of the feedback interaction between macro- and 
		microsystems.} 
		\label{KuramotoSynchronization}
		\end{figure}

Driven by dissipative forces, certain categories of systems are able to spontaneously develop long-range order all by themselves without 
the need of periodic force fields. This self-organization is a very general phenomenon. Famous examples are populations of chorusing 
crickets or blinking fireflies, cardiac pacemaker cells, laser arrays or coupled arrays of Josephson junctions, rhythmic applause or 
simply Huygens pendulum clocks [see Fig.~\ref{KuramotoSynchronization}(a)], which synchronize their oscillations when attached to the 
same wall \cite{Strogatz01}. Self-synchronization is generally driven by the feedback interaction between a macrosystem and microsystems 
[see Fig.~\ref{KuramotoSynchronization}(b)]. A macroscopic order parameter defines the boundary conditions for microscopic processes. 
On the other hand, these microscopic processes can, if they act collectively, influence the order parameter. This global feedback can 
give rise to instabilities and to long-range order. In the case of CARL the order parameter is a bunching of atoms, the microscopic 
processes are light scattering events, and the long-range order shows up as lasing accompanied by a simultaneous formation of an atomic 
lattice \cite{Courteille04}. 

The observations presented in the previous sections support this interpretation and confirm that collective interatomic coupling plays a 
prominent role in the dynamics of the atoms and the cavity modes. In the following sections we will briefly address some theoretical 
approaches to describe the dynamics quantitatively.

\subsection{Analytic treatment for perfect bunching}
\label{SecKuramotoAnalytic}

The collective dynamics of atoms simultaneously coupled to light modes with and without the presence of a ring cavity has been studied 
by Bonifacio \emph{et al.} \cite{Bonifacio95}, Gangl \emph{et al.} \cite{Gangl00} and Perrin \emph{et al.} \cite{Perrin02}. Our 
experiments deal with thermal clouds, so that the atomic coordinates may be treated classically. $x_n$ denotes the position of the 
$n^{th}$ atom and $p_n$ its momentum. Furthermore, due to its large mode volume, the cavity involves many photons in the dynamics, which 
allows us to describe the counterpropagating optical fields $\alpha_\pm$ classically as well. The only remaining quantization is that 
of the internal atomic excitation, whose dynamics essentially follows two-level Bloch equations. However, for the very large detunings 
used in our experiments the internal degrees of freedom can be adiabatically eliminated \cite{Gangl00,Perrin02,Bonifacio95}: At large 
detunings the field amplitudes and the atomic coordinates evolve much slower than the atomic excitation. Therefore, the time scales for 
the external and internal dynamics can be separated, which means that the atomic excitation is always in a (quasi-)stationary state and 
drops out of the equations of motion. 

Furthermore, large detunings allow us to neglect any contribution from radiation pressure, $\gamma_0\ll U_0$. The interaction of the 
atoms with the ring cavity is then governed by the following set of equations of motion \cite{Gangl00},\footnote{
	Friction forces due to optical molasses are not considered here.}
	\begin{align}
	& \dot{\alpha}_{\pm} =\left(-\kappa-iNU_0+i\Delta_c\right)\alpha_{\pm}-iU_0\sum\nolimits_n e^{\mp2ikx_n}\alpha_{\mp}
		+\eta_{\pm}~,\tag{7a}\nonumber\\
	& m\ddot{x}_n =2\hbar kiU_0\left(\alpha_+^{\ast}\alpha_-e^{-2ikx_n}-\alpha_+\alpha_-^{\ast}e^{2ikx_n}\right)~,\tag{7b}\nonumber
	\end{align}
with $\eta_-=0$. The equations have the following meaning: The first equation describes the modification of the field amplitudes by 
decay $\kappa$, detuning $\Delta_c$ of the laser from the cavity resonance, or collective light shift $NU_0$ (first term), by mutual 
scattering between the modes (second term) and external pumping (last term). Note that the mutual scattering amplitude is weighted with 
the location of the atoms modulo $2\pi/k$, i.e.~the photon scattering balance depends on the atomic locations. The second equation states 
that the dipole force sensed by the atoms is caused by the scattering of photons between the counterpropagating modes. 

The equations~(7) must, in general, be solved numerically. Analytic solutions are only available for special cases. For example, under 
the assumption of perfect atomic bunching, $e^{ikx_m}=e^{ikx_n}$ for all $m,n$, we can then replace all the individual atomic coordinates 
by a single pair of variables $x$ and $\dot{x}$. 

\bigskip

As discussed by Gangl \emph{et al.} \cite{Gangl00b} the dynamics of the CARL system critically depends on the detuning of the pump laser. 
An important point arising from experimental constraints is now the following: The pump mode $\alpha_+$ is tightly phase-locked to the 
cavity. As discussed in Ref.~\cite{Elsasser03}, the presence of atoms inside the cavity mode volume may shift the cavity resonance, and 
this shift depends on the atomic bunching. Since the phase-lock continuously works to readjust the detuning $\Delta_c$ (defined for the 
empty cavity) to the resonance (of the cavity filled with atoms), the detuning becomes a time-dependent variable, which follows the 
dynamics of the system. This fact needs to be considered by theories modeling our experiment, when it is operated in a parameter regime 
where the shift of the cavity resonances caused by the atoms, which is on the order of $NU_0$, exceeds the cavity linewidth $\kappa$. 
A good approach to model the impact of the phase-locking is to simply assume that the \emph{phase} of the pump laser $\eta_+$ is equal to 
that of the locked mode $\alpha_+$. This assumption is good, because the bandwidth of our servo system largely exceeds all characteristic 
frequencies of the coupled dynamics. Without loss of generality we may fix these phases to zero, while the phase $\phi$ of the probe 
mode remains a dynamic variable, 
	\begin{equation}
	\eta_+=|\eta_+| \qquad , \qquad \alpha_+=|\alpha_+| \qquad , \qquad \alpha_-=|\alpha_-|~e^{i\phi}~.\tag{8}\nonumber
	\end{equation}

The Eqs.~(7) can approximately be solved, if we assume that the field amplitudes have no explicit time-dependence, $\dot{\alpha}_\pm=0$. 
This assumption corresponds to an adiabatic elimination of inertial terms, which means that the standing wave and the atomic grating keep 
a fixed relative phase. It allows us to derive approximate solutions for the field amplitudes valid at long times, $t\gg\kappa^{-1}$ 
\cite{Cube04,Kruse03,Kruse04,Cube05}, 
	\begin{align}
	\alpha_+ & \approx \frac{\eta_+}{\kappa}~,\tag{9a}\nonumber\\
	|\alpha_-|^3 & \approx \frac{N^2U_0\eta_+}{24\kappa^2\epsilon t}~,\tag{9b}\nonumber\\
	(kv)^3 & \approx \frac{3\epsilon NU_0^2\eta_+^2}{\kappa}t~,\tag{9c}\nonumber
	\end{align}
where we introduced the abbreviation $\varepsilon\equiv\hbar k^2/m$. These formulae have been fitted to the data in 
Fig.~\ref{CollectiveAccelerate}(b) and (c).\footnote{
	It is interesting to compare the probe power dependence on the atom number $P_-\propto |\alpha_-|^2\propto N^{4/3}$ to the dependence 
	expected for superradiance $P_-\propto N^2$ (see Ref.~\cite{Robb04}).}

\subsubsection{Radiation pressure}
\label{SecKuramotoAnalyticPressure}

Far from atomic resonances, $\Delta_{D1}\approx-2\times10^5\Gamma$, the spontaneous scattering rate is many orders of magnitude lower 
than the coherent scattering rate $\gamma_0/U_0\simeq\Gamma/\Delta_{D1}\approx5\times10^{-6}$. On the other hand, the photon flux of the 
pump beam is quite large. When we compare the radiation pressure force to the CARL force resulting from Eq.~(7b), 
	\begin{align}
	F_{rp} & =2\hbar k\gamma_0|\alpha_+|^2~,\tag{10a}\label{Eq10a}\nonumber\\
	F_{carl} & <4\hbar kU_0|\alpha_+||\alpha_-|~,\tag{10b}\label{Eq10b}\nonumber
	\end{align}
we find that a power in the probe mode as low as $P_{probe}>P_{pump}~(\gamma_0/2U_0)^2$ is sufficient to ensure that the CARL force 
dominates. This shows that the radiation pressure can play a substantial role only at very long times, when the CARL acceleration has 
driven the probe mode very far out of resonance. But it does not take part in the acceleration process observed in 
Fig.~\ref{CollectiveAccelerate}(c). While the radiation pressure exerts a \emph{constant acceleration} on the atoms, we observe in 
experiment a \emph{cubic root time-dependence} of the CARL frequency.\footnote{
	Note that poor atomic bunching can also reduce the CARL force up to a point where radiation pressure overtakes.}

\subsubsection{Phase-locking by imperfect mirrors}
\label{SecKuramotoAnalyticPhaselocking}

Phase-locking is a known problem of ring laser gyroscopes \cite{Chow85}. Any impurity located on the mirror surfaces, such as small dust 
particles, may scatter light into the reverse mode. Hence, a standing wave builds up, which adjusts its phase such that the scattering 
rate and thus the energy transfer into the reverse mode is maximized. The phase of the standing wave thus locks to the impurity. 

The dipole force resulting from this standing wave can substantially alter the dynamics of atoms interacting with the cavity. In the 
worst case, the CARL acceleration is entirely inhibited, because the atoms are trapped in the standing wave caused by the mirror 
backscattering. In our ring cavity the phase fronts of the light fields hit the mirrors at angles of $22.5^\circ$ or $45^\circ$. 
Macroscopic dust particles on the mirror surfaces traverse many phase fronts, so that in the average over several standing wave periods 
the net scattering between the modes should vanish. 

However, the high finesse of the cavity amplifies even very small net scattering imbalances induced by microscopic irregularities 
(spatial noise). To see this, we describe the imbalance as originating from a distribution of $N_s$ point-like scatterers \emph{fixed} 
at locations $x_n$. With the scattering efficiency $U_s$ and assuming for simplicity $\Delta_c=NU_s$, the stationary solution of 
Eq.~(7a) reads 
	\begin{equation}
	\frac{\alpha_-}{\alpha_+}=-i\frac{N_sU_sb_s}{\kappa}~.\tag{11}\nonumber
	\end{equation}
where $b_s\equiv N_s^{-1}\sum_n e^{2ikx_n}$. This equation shows that, at first sight counterintuitively, the intensity in the reverse 
mode increases, when the finesse of the mirrors is improved (or $\kappa$ is reduced). The fraction measured in experiment, 
$|\alpha_-|^2/|\alpha_+|^2\approx 0.35\%$, corresponds to a collective coupling strength of about $N_sU_s\approx 0.06\kappa$, which is 
smaller but in the same order of magnitude as the atom-field coupling used for observing CARL activity. 
 
Furthermore, inserting the solution~(11) into the expression for the dipole force~(7b), we find that the phase of the backscattered 
light adjusts in such a way that the scatterer sits at half height on the edge of the resulting standing wave. In this configuration the 
backscattering rate is largest. However, any spatial extend of the scatterer's size, expressed by a decrease of the bunching $b_s$, will 
diminish this rate. 
		\begin{figure}[h]
		\centerline{\scalebox{0.5}{\includegraphics{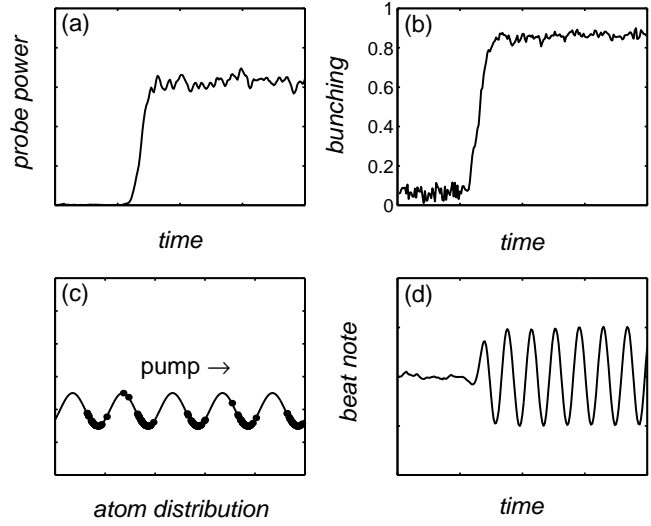}}}\caption{
		Simulation of CARL action triggered by suddenly turning on  an optical molasses. The figures show \textbf{(a)} the evolution of 
		the probe photon 	number in the reverse cavity mode, \textbf{(b)} the bunching of the atoms, \textbf{(c)} the atomic density 
		distribution along the cavity axis at the end of the simulation, and \textbf{(d)} the beat signal between the cavity modes.}
		\label{KuramotoLangevin}
		\end{figure}

\subsection{Simulations of atomic trajectories with friction and diffusion}
\label{SecKuramotoTrajectories}

The CARL equations~(7) predict unlimited acceleration of the atoms and of the phase of the standing wave formed by the pump mode and the 
reverse mode. To balance this acceleration, we have introduced in our experiment an optical molasses. As discussed in 
Sec.~\ref{SecCollectiveExperimentMolasses}, at low atomic velocities molasses are well characterized by a velocity-independent friction 
coefficient \cite{Dalibard89}. For the theoretical description we may thus supplement Eq.~(7b) by a friction force 
$\gamma_{fr}\dot{x}_n$. 

Through the friction coefficient, the frequency and the amplitude of the probe beam depend on the settings of the molasses. This is 
verified experimentally \cite{Cube04}. However another prediction of this model is not verified. Simulations of Eqs.~(7) 
including friction show a complete synchronization of the atomic trajectories for any set of parameters. After an amount of time, which 
depends on coupling strength, atom number and pump power, the atomic bunching tends to unity. This is due to the absence of any influence 
working to counteract the atomic ordering from the model, and contradicts the observation of a threshold observed in experiment 
\cite{Cube04}. 

\bigskip

A heating mechanism which is well-known to occur in optical molasses is momentum-diffusion. Momentum transfer to the atoms due to 
photonic recoil by scattering of light leads to a random walk of the atoms, which can be described as resulting from a random force 
\cite{Robb04}. Therefore, we describe the dynamics of our system by a set of Langevin equations. In general these cannot be solved 
analytically, and we have to numerically iterate them in time. For the sake of computational efficiency, we only calculate the 
trajectories of $100$ atoms. The molasses friction is strong enough to completely overdamp the atomic motion, so that we may 
adiabatically eliminate the atomic inertia and write \cite{Robb04}, 
	\begin{align}
	\dot{\alpha}_{\pm} & =-\left(\kappa+iNU_0-i\Delta_c\right)\alpha_{\pm}-iU_0\sum\nolimits_n e^{\mp2ikx_n}\alpha_{\mp}
		+\eta_{\pm}~,\tag{12a}\nonumber\\
	0 & =2\hbar kiU_0\left(\alpha_+^{\ast}\alpha_-e^{-2ikx_n}-\alpha_+\alpha_-^{\ast}e^{2ikx_n}\right)-\gamma_{fr}\dot{x_n}
		+F_{noise}(t)~.\tag{12b}\nonumber
	\end{align}

For the simulation, we assume that the atoms are irradiated by a pump laser and that there is initially \emph{one} photon in the probe 
mode. We start with a homogeneous atomic cloud distributed around the waist of the cavity mode and having a thermal momentum 
distribution. Fig.~\ref{KuramotoLangevin} shows the time evolutions of the backscattered light intensity $|\alpha_-|^2$, the bunching 
parameter\footnote{
	The bunching parameter measures the periodic arrangement of the atoms. It is also called the order parameter or Debye-Waller factor.} 
$b\equiv N^{-1}\left|\sum\nolimits_n e^{2ikx_n}\right|$, the spatial distribution $x_n$ of the atoms, and the interference signal 
$|\alpha_++\alpha_-|^2$. A short time delay after the optical molasses is switched on, the amplitude of the backscattered beam and the 
order parameter suddenly start to grow towards macroscopic values. The atoms arrange themselves into a periodic lattice, synchronize 
their velocities and, together with the standing wave, propagate in the pump beam direction [arrow in Fig.~\ref{KuramotoLangevin}(c)]. 
In the same time, the beat signal starts to oscillate. The calculated trajectories reproduce quantitatively the experimental 
observations shown in Figs.~\ref{CollectiveObservation} to \ref{CollectiveTof}. 
		\begin{figure}[h]
		\centerline{\scalebox{0.5}{\includegraphics{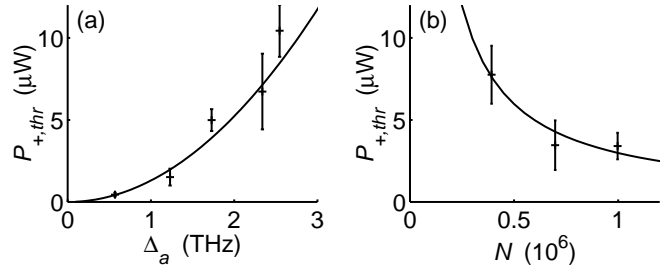}}}\caption{
		Behavior of the pump power threshold $P_{+,thr}$ when \textbf{(a)} the detuning $\Delta_a$ or \textbf{(b)} the atom number $N$ is 
		varied. The fits are obtained from a linear stability analysis of a Fokker-Planck equation \cite{Cube04,Robb04}.}
		\label{KuramotoThreshold}
		\end{figure}

\subsubsection{Lasing threshold}
\label{SecKuramotoTrajectoriesThreshold}

The simulations confirm that, under the influence of an optical molasses and mediated by collective atomic recoil, the coupled 
atom-cavity system generates laser light. In the same time, the atoms self-organize into a periodic lattice, while breaking translation 
symmetry. The sudden appearance of a coherent radiation reminds a laser threshold, and the atomic ordering corresponds to a 
thermodynamic phase transition. To model the phase transition we have chosen an alternative approach. Instead of simulating single-atom 
trajectories, we derive from the Eqs.~(12) a Fokker-Planck equation for the atomic density distribution. The solution of this 
equation reproduces the time evolution of the density distribution, and a linear stability analysis leads to simple analytic formulae 
describing the threshold conditions \cite{Robb04,Javaloyes04}. 

The calculated threshold behavior is observed in experiment: To trigger the CARL and obtain probe laser emission, a minimum pump laser 
power is needed. Below this threshold the order parameter disappears, far above threshold it approaches unity. The threshold power 
should obviously depend on some control parameters. Fig.~\ref{KuramotoThreshold} shows that, as expected, the threshold increases when 
the coupling strength is decreased, i.e.~if the pump laser is tuned further away from the atomic resonance \cite{Cube04}. The threshold 
decreases when the atom number is increased. 

The role of the molasses in the CARL dynamics is twofold: On one hand, the dissipation of energy permits to reach a steady-state. On the 
other hand, atomic momentum-diffusion processes, which are intrinsically connected to optical molasses, limit the equilibrium 
temperature. The interplay of dissipation and diffusion rules the thermodynamic phase transition, and gives rise to the observed 
threshold behavior of the CARL radiation.

\subsubsection{Self-synchronization}
\label{SecKuramotoTrajectoriesSynchronization}

The simplest model describing synchronization phenomena, known as Kuramoto model \cite{Kuramoto84}, considers ensembles of coupled limit 
cycle oscillators. It predicts the occurrence of spontaneous synchronization, when a critical number of oscillators is put together or 
when the coupling strength exceeds a critical value \cite{Strogatz00}. Under certain approximations, the Kuramoto model is also 
applicable to describe the CARL bunching of atomic trajectories \cite{Cube04}. However in the case of CARL, unlike for the Kuramoto 
model, the collective oscillation frequency is self-determined. The observed probe mode emits a \emph{new} laser frequency, which only 
depends on \emph{global} system parameters. 
		\begin{figure}[h]
		\includegraphics[width=5.4cm]{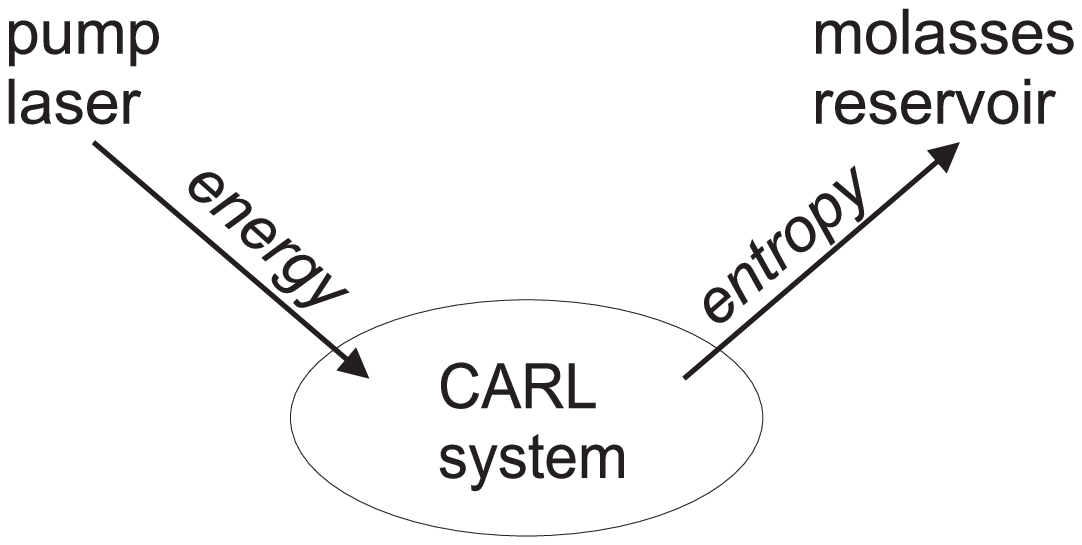}~(a)
		\hfil
		\includegraphics[width=8.5cm]{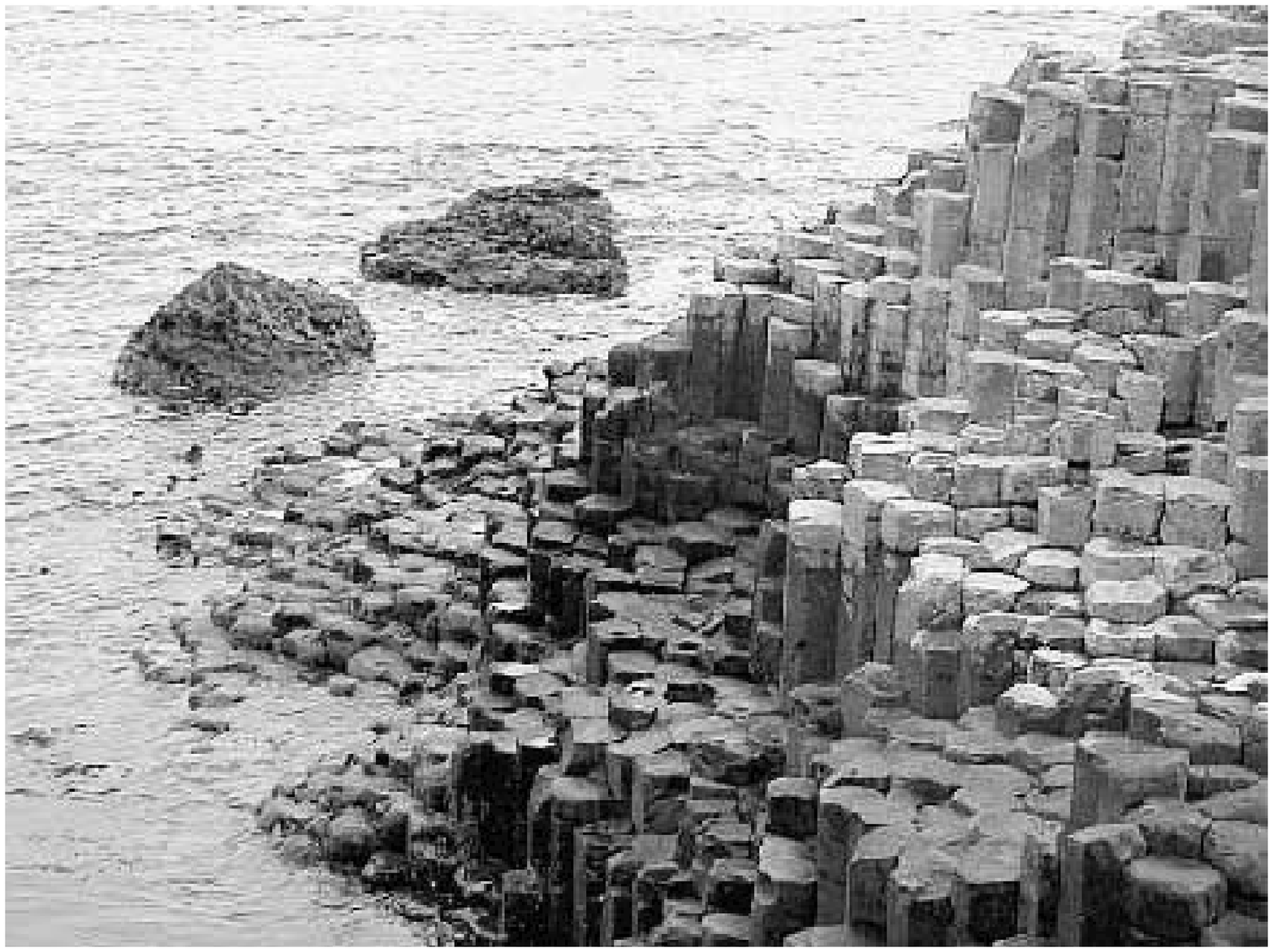}~(b)
		\caption{\textbf{(a)} Scheme of the CARL coupling to the environment. \textbf{(b)} Self-organization at Giant's Causeway in 
		Northern Ireland. The hexagonal rock formations are frozen B\'enard convection flow patterns.} 
		\label{KuramotoInterface}
		\end{figure}

In fact, CARL represents in many respects an ideal system to study self-synchronization phenomena: First of all, free atoms are very 
simple systems, the behavior of which is almost completely understood. They are among the smallest entities, which can be handled and 
controlled with current technologies. Therefore, many of them can be joined together, thus experimentally realizing the thermodynamic 
limit to a good approximation. Furthermore, a truly universal coupling force (i.e.~a force which couples any pair of atoms with the same 
strength, independently of their distance) is desirable, because the system is much easier to understand if local peculiarities do not 
affect the global dynamics. The CARL coupling is universal, because it is mediated by a light mode which is completely delocalized 
within the ring cavity. Finally, the coupling force should be sufficiently strong, so that the subsystems do not evolve independently 
from each other but execute collective actions. On the other hand, the coupling to the external world, e.g.~reservoirs or pumps should 
be sufficiently weak to allow the system to develop a self-determined behavior. 

The atomic self-synchronization corresponds to a thermodynamic phase transition, where the control parameter is time. It is ruled by the 
competition between the dynamical coupling which generates order, and diffusion which is a source of disorder. It is however important 
to note that the system is never in thermal equilibrium. To maintain order, one has to constantly inject energy via the pump laser [see 
Fig.~\ref{KuramotoInterface}(a)]. The produced entropy is dissipated via the reservoir of the optical molasses. This introduces the 
irreversibility required in any lasing process. The transition towards a structure with a higher symmetry has certain analogies with the 
dissipative structures postulated by Prigogine \cite{Nicolis77} or with Haken's synergetic systems \cite{Haken75}, whose most famous 
examples are the B\'enard structures. Those are regular patterns of convection flow arising in a heated fluid, and can e.g.~be observed 
as frozen geological rock formations [see Fig.~\ref{KuramotoInterface}(b)].

\section{Probing long-range order}
\label{SecBragg}

The hallmark of collective atomic recoil lasing is atomic bunching. The CARL signatures presented in Sec.~\ref{SecCollectiveSignatures} 
are more or less indirect indications for the presence of bunching. There is however a way to probe atomic gratings directly, which is 
Bragg scattering. 
		\begin{figure}[ht]
		\centerline{\scalebox{0.5}{\includegraphics{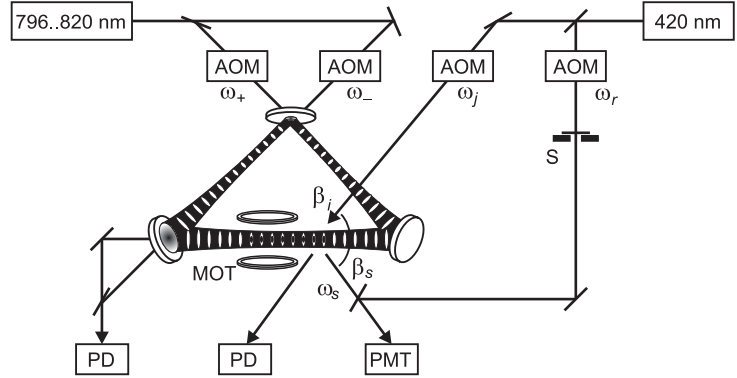}}}\caption{
		Modified setup for Bragg spectroscopy. The ring cavity is now pumped in \emph{both} directions with frequencies $\omega_\pm$ 
		tuned by means of acousto-optic modulators (AOM). Probe light, $\omega_j$, is shone under the Bragg angle onto the atoms. The 
		transmitted light is monitored with a photodetector (PD), Bragg reflection is detected with a photomultiplier (PMT). By opening a 
		shutter (S) the Bragg-reflected beam can be heterodyned with a reference frequency $\omega_r$.}
		\label{BraggBeatsetup}
		\end{figure}

Bragg scattering at three-dimensional near-resonant gaseous atomic gratings (called optical lattices) has been demonstrated 10 years ago 
by Birkl \emph{et al.} and Weidem\"uller \emph{et al.} \cite{Birkl95,Weidemuller95,Weidemuller98}. In contrast to those experiments, our 
sample consists of an \emph{one-dimensional} array of atomic clouds confined at the antinodes of a ring cavity standing wave, and the 
standing wave light is tuned very \emph{far from atomic resonances}, so that the confining lattice potential is conservative and 
dissipative cooling forces are absent \cite{Slama05}. 

First attempts to employ the Bragg scattering method to monitor CARL bunching have not been successful. On one hand, this is due to the 
presence of atomic gratings \emph{not} related to CARL. Spurious mirror backscattering (see Sec.~\ref{SecKuramotoAnalyticPhaselocking}) 
generates a stationary standing wave fraction, which is not generated by CARL dynamics of trapped atoms, but nevertheless forces the 
atoms into a partial alignment. To solve this problem, having in mind that CARL-bunched atoms always propagate along the cavity axis, we 
have developed a method of detecting \emph{moving} lattices. On the other hand, the weak Bragg scattering efficiency of typically 
$R=0.1\%$ makes the detection of bunching a challenge. Fortunately, we have been able to show that suitable modifications of the 
experiment can dramatically increase the Bragg reflection \cite{Slama05c} up to a point, where it should clearly be sufficient to detect 
CARL-bunching in future experiments.

\subsection{Bragg scattering}
\label{SecBraggScattering}

The setup for Bragg scattering is shown in Fig.~\ref{BraggBeatsetup}, with the shutter S closed. We test the setup by generating an 
atomic lattice by imposing a periodic force field. To this end both counterpropagating modes of the ring cavity are pumped with the 
same light frequency, $\omega_+=\omega_-$, so that the atoms arrange themselves in an optical grating. To probe the grating, light at 
$420~$nm resonant to the 5$S_{1/2}$-6$P_{3/2}$ transition is irradiated under the Bragg angle $58^{\circ}$, and the reflected signal is 
recorded. 
		\begin{figure}[ht]
		\centerline{\scalebox{0.85}{\includegraphics{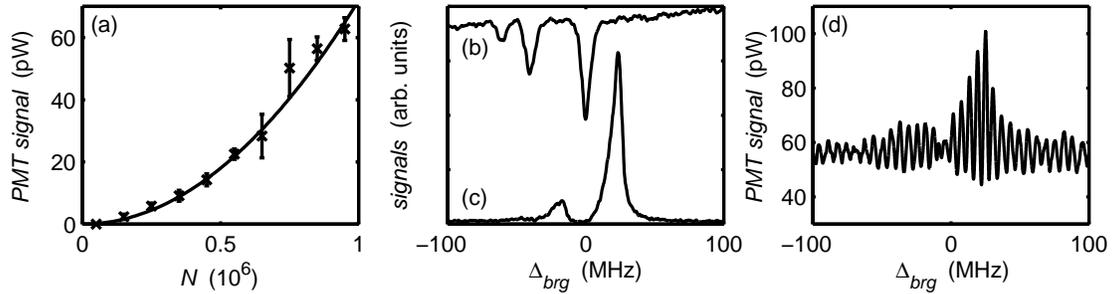}}}\caption{
		\textbf{(a)} Dependence of the Bragg scattered peak intensity from the atom number \cite{Slama05}. 
		\textbf{(b)} Absorptionspectrum from a cloud confined in a magneto-optical trap (for reference).  
		\textbf{(c)} Bragg reflection spectrum. 
		\textbf{(d)} Spectrum of the beat between the Bragg-reflected light and the incident mode.}
		\label{BraggAtomnumber}
		\end{figure}

There are many obvious signatures of Bragg scattering: First of all, Bragg scattering works only for a well-defined angle of incidence. 
In fact the observed acceptance angle and the beam divergence of the reflected beam are both on the order of $0.1^{\circ}$. Second, the 
reflected intensity should scale with the square of the atom number and not linearly, as one would expect for isotropic resonance 
fluorescence \cite{Slama05}. This is verified in Fig.\ref{BraggAtomnumber}(a). Fig.\ref{BraggAtomnumber}(b) shows a transmission 
spectrum of magneto-optically trapped atoms, which serves as a frequency reference. An example of a Bragg reflection spectrum is shown 
in Fig.~\ref{BraggAtomnumber}(c). The spectrum is displaced due to the light-shift of the atoms in the dipole trap. 

The trapped atoms sense a position-dependent dynamical Stark-shift, which results in inhomogeneous broadening of the atomic transitions. 
This broadening and also atomic heating induced by the Bragg laser limited the Bragg reflection efficiency in early experiments 
\cite{Slama05}. Subsequent experiments have shown that the Bragg scattering efficiency can be dramatically improved by tuning the probe 
laser to the $D_2$ line at $780~$nm. Efficiencies up to $R=30\%$ have been reached allowing for studies of subtle effects arising from 
the one-dimensional geometry of the optical lattice \cite{Slama05b}. We have even observed signatures of a tricky interplay between 
multiple reflections and diffuse scattering in thick lattices \cite{Slama05c}.

\subsection{Heterodyned Bragg spectra}
\label{SecBraggBeating}

The Bragg signals have been obtained on atoms confined in a standing wave, i.e.~with bidirectional pumping of the ring cavity. The Bragg 
scattering technique can be refined to be sensitive to \emph{moving} long-range order by heterodyning the Bragg-scattered light with a 
reference \cite{Slama05}. We test the idea by simulating the CARL behavior, i.e.~we create a rotating standing wave inside the ring 
cavity by supplying \emph{different pump frequencies} for the counterpropagating modes, $\omega_+\neq\omega_-$. The frequencies are 
shifted by means of two acousto-optic modulators (AOM). The Bragg-reflected beam is phase-matched with a reference laser beam [the 
shutter (S) in Fig.~\ref{BraggBeatsetup} is opened], and the frequency beat is monitored on a photomultiplier (PMT). By means of another 
AOM the frequency of the reference beam $\omega_r$ is chosen such that the beat occurs at frequencies low enough to be resolved by our 
PMT. A fourth AOM ramps the Bragg beam frequency $\omega_j$ across resonance (see Fig.~\ref{BraggBeatsetup}). 

\begin{figure}[ht]
		\centerline{\scalebox{0.7}{\includegraphics{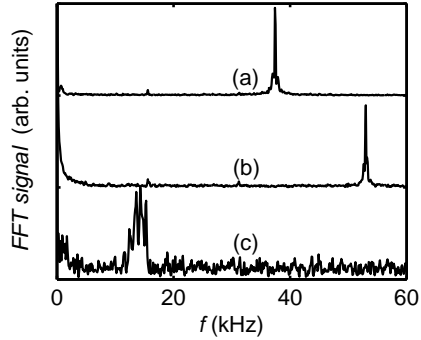}}}\caption{
		Detection of moving Bragg lattices. Shown are the Fourier spectra of \textbf{(a)} $2k_{\mathrm{dip}}v=37~$kHz, \textbf{(b)} 
		$\Delta\omega_j=53~$kHz and \textbf{(c)} $\Delta\omega_s\approx16~$kHz.}
		\label{BraggMoving}
		\end{figure}

The observed beat signal between Bragg and reference beam is shown in Fig.~\ref{BraggAtomnumber}(d). To evaluate the beat frequency, we 
show in Fig.~\ref{BraggMoving} three Fourier spectra: (a) The beat of the counterpropagating cavity modes oscillates at the frequency 
$2k_{\mathrm{dip}}v$; (b) a reference interferometer (not shown in Fig.~\ref{BraggBeatsetup}) measures the difference frequency 
$\Delta\omega_j=\omega_j-\omega_r$, and (c) the frequency of the Bragg beat is $\Delta\omega_s=\omega_s-\omega_r$. The fact that the 
Bragg-scattered light is obviously frequency-shifted by an amount corresponding to the Doppler-shift of the moving grating, 
	\begin{equation}
	\Delta\omega_s=\Delta\omega_j-2k_{\mathrm{dip}}v~,\tag{13}\nonumber
	\end{equation}
permits to determine the lattice propagation velocity $v$ with an accuracy better than 1\% by measuring $\Delta\omega_j$ and 
$\Delta\omega_s$. 

Retardation effects resulting from unequal lengths of the interferometer arms lead to shifts in the recorded beat frequencies, as the 
laser frequency is scanned across resonance. The shifts can be corrected by inverting the scan direction. Furthermore, the Fourier 
spectrum of the Bragg beat in Fig.~\ref{BraggMoving}(c) shows asymmetric sidebands. These are artifacts arising from the presence of two 
hfs-resonances joint to the fact that the signal is amplitude- and phase-modulated by the frequency scan. 

\subsection{Measuring the Bragg scattering phase}
\label{SecBraggOutlook}

As shown in the previous section, heterodyned Bragg scattering is well-suited not only to probe long-range order, but also to detect 
its motional dynamics. Hence, it constitutes an interesting tool to study the dynamics of CARL. But there is more information in the 
heterodyne signal, because it gives access to the scattering phase. 

To see this, let us write the global response of the atomic cloud to an incident laser $E_j=E_{j0}e^{i\omega_j t}$, as a complex 
scattering amplitude $r=Ae^{i\phi}$. The Bragg-scattered light amplitude is then given by $E_s=r E_j=E_{j0}A(t)e^{i\omega_j t+i\phi(t)}$. 
Heterodyning the Bragg-scattered light with a reference beam, $E_r=E_{r0}e^{i\omega_r t}$, yields the signal 
	\begin{equation}
	S\propto|E_r+E_s|^2\approx E_{j0}^2+2rE_{r0}E_{j0}\cos\left[(\omega_j-\omega_r)t+\phi\right]~,\tag{14}\nonumber
	\end{equation}
if $rE_{j0}\ll E_{r0}$. The amplitude and phase information are extracted by demodulating the beat signal with the reference frequency 
$\Delta\omega_j$ using two different phases. From these quadrature components we have determined the amplitude profile $A(\Delta)$ and 
the phase profile $\phi(\Delta)$ of the resonance \cite{Slama05}. The amplitude profile corresponds to the spectrum observed earlier by 
standard Bragg scattering. But the phase profile represents the first direct measurement of the phase-lag due to elastic scattering of 
light at an atomic cloud. 

Under normal conditions phase measurements are not easy, because in thermal clouds Rayleigh scattering is Doppler-broadened by photonic 
recoil. In our case, the atoms are not only cold, but strongly confined; the Lamb-Dicke factor is about 100. Thus velocity-dependent 
recoil shifts are totally absent. Furthermore, Rayleigh scattering is normally weak (too weak to be heterodyned) because the atoms must 
be pumped at very low powers to avoid inelastic resonance fluorescence. Therefore, we gain a lot by resonantly enhancing the elastic 
peak by arranging the atoms into a grating and scattering at exactly the Bragg angle. To resume, two conditions are helpful for 
measurements of Rayleigh scattering phases, \emph{strong atomic confinement} and \emph{atomic ordering}.

\section{Conclusion}
\label{SecConclusion}

With our work, we have shown that atoms interacting with the counterpropagating modes of a high-finesse ring cavity can exhibit 
collective dynamics, provided the interatomic coupling constant is sufficiently large as compared to the cavity linewidth. An important 
prerequisite for the possible use of ring cavities for quantum information processing \cite{Hemmerich99} is thus experimentally verified. 
The epitome of collective behavior is without doubts the collective atomic recoil laser, the signatures of which we have detected for the 
first time. Previous attempts to observe CARL \cite{Lippi96b,Hemmer96} were based on a maximization of the coupling strength by tuning 
the pump laser close to an atomic resonance. In contrast, the use of a high-finesse ring cavity allowed us to obtain strong coupling at 
huge pump laser detunings, where no noticeable effect is expected from atomic polarization gratings which do not correlate to density 
gratings and where spontaneous scattering plays no role. 

Another important result of our work is the proof that CARL can be made stationary. Optical molasses not only introduce very efficient 
dissipation forces, but also a coupling to a finite temperature reservoir giving rise to spontaneous self-synchronization beyond 
a critical pump power or coupling strength. Just like self-synchronization, which is a very common phenomenon known to give rise to 
macroscopic consequences such as large scale pattern formation, CARL is a purely classical phenomenon. It could in principle be observed 
with billiard balls, if there was a suitable coupling force at hand \cite{Wiggins02}. On the other hand, our system consists of atoms, 
which are microscopic particles. Therefore it ought to exhibit quantum features in some circumstances, in particular at ultra-low 
temperatures, where quantum statistics come into play. E.g.~we may expect effects like quantized acceleration \cite{Piovella01b}, 
superradiance \cite{Piovella01} or quantum entanglement \cite{Moore99b,Piovella03}. 

The study of the quantum regime will be the challenge of our future research efforts. We plan to cool the atoms to very low 
temperatures, eventually even beyond the limit to quantum degeneracy. It is clear that optical molasses cannot be used to exert friction 
forces on Bose-Einstein condensates. However there are other dissipation mechanisms working in cavities, such as cavity-cooling 
\cite{Vuletic00}. This cooling mechanism is ultimately a consequence of the coupling of the cavity to a zero-temperature reservoir via 
the finite transmission of the mirrors; vacuum fluctuations couple to the optical cavity, while thermal photons are frozen out, which 
corresponds to a situation of having dissipation without diffusion. If a threshold behavior is observed under such circumstances, it is 
an interesting question to ask whether it is linked to a quantum phase transition. In practice, a controllable dissipation mechanism 
working for Bose-Einstein condensates would be an extremely valuable tool, especially for the study of decoherence mechanisms. 
Superfluidity makes the condensates insensitive to collisions, and other sources for dissipation are not available. 

Regarding quantum information applications, the aptitude of ring cavities is conditionned to a totally coherent evolution of the atomic 
states and the absence of uncontrolled decoherence. This means that thermal motion must be completely avoided. The sequel of the ring 
cavity experiment in the ultralow temperature regime thus seems compulsory.

\begin{acknowledgement}
We are grateful for valuable discussions with R. Bonifacio, J. Javaloyes, R. Kaiser, G.-L. Lippi, M. Perrin, N. Piovella, A. Politi, 
H. Ritsch, and G.R.M. Robb. We acknowledge financial support from the Landes\-stiftung Baden-W\"urttemberg. 
\end{acknowledgement}

\bigskip

\bibliographystyle{Carl}
\bibliography{D:/Arbeit/Organisieren/papers}

\end{document}